# Intercalation of a Nonionic Surfactant ($C_{10}E_3$) bilayer into a Na-Montmorillonite Clay

*Régis Guégan*[*]

Institut des Sciences de la Terre d'Orléans, CNRS-Université d'Orléans, 1A Rue de la Ferollerie, 45071 Orléans Cedex 2, France

regis.guegan@univ-orleans.fr

**November 09, 2010**

**[*]**To whom correspondence should be addressed. E-mail: regis.guegan@univ-orleans.fr. Phone: +33 (0) 2 38 49 25 41. Fax: +33 (0) 2 38 63 64 88

ABSTRACT

A nonionic surfactant, the tri-ethylene glycol mono n-decyl ether ($C_{10}E_3$), characterized by its lamellar phase state, was introduced in the interlayer of a Na-montmorillonite clay at several concentrations. The synthesized organoclays were characterized by Small Angle X-Ray Scattering in conjunction with Fourier Transform Infrared spectroscopy, and adsorption isotherms. Experiments showed that a bilayer of $C_{10}E_3$ was intercalated into the interlayer space of the naturally exchanged Na-montmorillonite, resulting in the aggregation of the lyotropic liquid crystal state in the lamellar phase. This behavior strongly differs from previous observations of confinement of nonionic surfactants in clays where the expansion of the interlayer space was limited to two monolayers parallel to the silicate surface and cationic surfactants in clays where the intercalation of organic compounds is introduced into the clay galleries through ion exchange. The confinement of a bilayer of $C_{10}E_3$ nonionic surfactant in clays offers new perspectives for the realization of hybrid nanomaterials since the synthesized organoclays preserve the electrostatic characteristics of the clays, thus allowing further ion exchange, while presenting at the same time a hydrophobic surface and a maximum opening of the interlayer space for the adsorption of neutral organic molecules of important size with functional properties.

# Introduction

Organoclays materials are extensively used in a wide range of applications as a basis for the implementation of polymer layered nanocomposites, new hybrid nanocomposites for photophysical applications, paints, cosmetics; refractory varnish; thixotropic fluids or geochemical barriers in waste landfills…[1-5] Indeed; organoclays combine the properties of natural clays such as a high surface area and a hydrophobic surface which allow the adsorption of neutral organic compounds or the dispersion of clays into polymers.[1,4,6] Another feature is the strong increase of the interlayer height, comparatively small for the clay before modification, which makes easier the insertion of guest molecules whose orientations follow these-ones of the previously adsorbed molecules used for the manufacture of organoclays.[6]

The synthesis of organoclays is based on the mechanisms of the reactions that the clay minerals can have with the organic compounds.[1,7] Organo-montmorillonites are mainly obtained by intercalating cationic surfactants such as quaternary ammonium compounds into the interlayer space through ion exchange.[8-11] Depending on the packing density of the alkylammonium ions, different arrangements of organic molecules can be formed.[9] The interlayer spacing of these organoclays generally increases as the surfactant loading increases and reaches a saturation limit which corresponds to or larger than the clay cation exchange capacity (CEC). However the intercalation of long alkyl tails surfactant within the interlayer space is irreversible and prevents any further cationic exchanges which limit, as a result, the potential applications of the nanocomposites and/or the innovation of new hybrid materials.

Several authors have raised the interest to realize organoclays by the use of other surfactants. Polyhedral oligomeric silsequioxane (POSS) imidazolium cationic surfactant is intercalated in clay by ion exchange.[12-14] Recent studies have shown a large interlayer space at low surfactant loadings far below the CEC which are related to the organization of the POSS molecules in a bilayer packing in the bulk phase. The relationship between the molecular structures of the POSS surfactants and the way they pack within the interlayer space is not completely understood.

Nonionic surfactants can self assemble to form bicontinuous structures made of amphiphilic bilayers as POSS surfactants can displayed in bulk phase.[15] The created structures can be flat into planar sheets (lamellar phase) or closed (vesicules, onions structures) which play an important role in biological or industrial processes as the use of templates for the realization of nanoporous materials.[16-18] Nonionic surfactant systems containing n-$C_nH_{2n+1}(OCH_2CH_2)_mOH$ surfactant (abbreviated as $C_nE_m$) were demonstrated to be sensitive to temperature changes. In presence of water, $C_nE_m$ shows, above the critical micelle concentration (CMC), polyatomic formed structures, leading to the existence of various liquid crystalline phase such as lamellar, cubic, sponge and hexagonal phases. The $C_{10}E_3$ (tri-ethylene glycol mono n-decyl ether) and water system forms internal structures organized by surfactant bilayers such as the lamellar ($L_\alpha$) and sponge ($L_3$) phases. The lamellar phase $L_\alpha$ is characterized by a stack of surfactant molecules aggregated into a two dimensional structure (membranes) and layers of water whose organization is ordered at long ranges and displays a smectic order.

The incorporation of nonionic surfactants into porous or layered materials is a non trivial task and the study of these hybrid materials can offer helpful answers concerning the use of such organoclays and their derivatives for possible applications in the case of waste landfills or nanocomposites fields.[19] Previous studies showed that nonionic surfactants adsorption usually is limited to a monolayer or for high loadings of surfactants to double molecular layers parallel to the silicate surface.[20-21] The adsorption of these neutral surfactants on the hydrophilic clay surface implies various chemical interactions (ion-dipole interaction, ion-dipole via intermediate hydrogen bonds, direct hydrogen bonds (H-bonds) with the surface, Van der Waals forces) and is still a subject of debate.[20-21]

This study shows the adsorption of a nonionic surfactant, the triethylene glycol monodecyl ether ($C_{10}E_3$), in a lamellar phase $L_\alpha$, on a Na-montmorillonite clay from Wyoming. The synthesized composites $C_{10}E_3$-clays were characterized by several complementary techniques: Small Angle X-Ray

Scattering (SAXS), adsorption isotherms and Fourier Transform Infrared Spectroscopy (FTIR). The performed experiments revealed a confinement of $C_{10}E_3$ in a bilayer within the interlayer space.

## Experimental Section

### Host Layered Materials

The montmorillonite (Na-smectite) used in this study was supplied by the Clay Minerals Society. This clay originates from the Newcastle formation (cretaceous), Crook County, Wyoming. The chemical composition of the montmorillonite is 62,9% $SiO_2$, 19.6% $Al_2O_3$, 3.35% $Fe_2O_3$, 3.05% MgO, 1.68% CaO, and 1.53% $Na_2O$. The formula of the montmorillonite can be expressed as $(Ca_{0.12}Na_{0.32}K_{0.05})[Al_{3.01}Fe(III)_{0.41}Mn_{0.01}Mg_{0.54}Ti_{0.02}][Si_{7.98}Al_{0.02}]O_{20}(OH)_4$, as calculated from its chemical composition.[4,5] The clay was fractioned to <2 µm by gravity sedimentation and purified by well-established procedures in clay science.[4,5] Sodium-exchanges were prepared by immersing the clay into 1 M solution of sodium chloride. Cation exchange was completed by washing and centrifuging four times with dilute aqueous NaCl. Samples were finally washed with distilled-deionized water and transferred into dialysis tubes to obtain chloride-free clays and then dried at room temperature. The cation exchange capacity (CEC) of the resulting Na-smectite is about 80.2 mequiv/100g.

### $C_{10}E_3$ nonionic surfactant

The surfactant $C_{10}E_3$ was purchased from Nikko Chemicals, Inc., (Tokyo, Japan) and used without further purification. Distilled and deionized water was used to prepare the aqueous solutions of surfactant. All samples were prepared by weight and the volume fraction of surfactant was calculated using the following density: 0.938 g/cm$^3$.[22] In this binary system, an isotropic sponge phase ($L_3$) phase exists at high temperature whereas a lamellar ($L_\alpha$) phase exists at low temperature as shown in Figure 1. The interlamellar spacings of the dilute lamellar phase can reach several hundreds nanometers and can be described, assuming conservation of volume and a constant bilayer thickness, $\zeta$, by eq (1):

$$d = \frac{\zeta}{\varphi} \quad (1)$$

where d is the repeat distance, φ the volume fraction of surfactant in solution and the $C_{10}E_3$ bilayer thickness ζ is assumed to be close to 27.6 Å.[15]

## Preparation of the $C_{10}E_3$ / clay composites

Several aqueous solutions of $C_{10}E_3$ were prepared at room temperature above the critical micellar concentration (CMC) which is $6.10^{-4}$ mol.L$^{-1}$ at concentrations varying from $3.3.10^{-3}$ to $1.3.10^{-2}$ mol.L$^{-1}$ thus displaying the lamellar phase ($L_α$) which was checked by phase contrast optical microscopy (insert in Figure 1).[15,22] Solutions were homogenized by stirring at room temperature for several days. Clay suspension solutions were then dispersed into the surfactant solutions. The average pH value of the solutions was about 6.5 ± 0.2 and remained constant during the synthesis. The suspensions obtained were then stirred for 24 h at 250 rpm. The organoclay residues, separated by centrifugation, were washed twice by water and dried at 70°C for 48 h before being crushed using an agate mortar.

## Adsorption isotherms

The concentration of carbon in solution were measured using an element analyzer (Shimadzu TOC 5050 /SSM 5000-A) and were carried out at the Service Central d'Analyse du CNRS at Solaize in France. Adsorbed surfactants amount were determined by the difference between the added surfactant and that which remained in the solution and allowed us to determine the equilibration isotherms. The concentration of carbon measured in the obtained resulting solution after the washing treatment of the organoclays was close to this-one of the blank water in the range of the standard deviation error. This last observation indicated that confined $C_{10}E_3$ molecules are in strong interaction with the silicate surface.

## Fourier Transform Infrared (FTIR) Spectroscopy

Infrared spectra, in the region of 550-4000 cm$^{-1}$, were determined by a Brucker Equinox 55S infrared spectrometer equipped with a diamond attenuated total reflection (ATR) and deuterated triglycine sulfate (DTGS) detector. Each spectrum was the average of 200 scans collected at 1 cm$^{-1}$ resolution. Samples were in the powder form and were deposited on diamond and pressed by an aluminum anvil.

**Small-angle X-ray scattering (SAXS)**

Scattering measurements were performed on the small-angle scattering instrument SWING at SOLEIL (Synchrotron SOLEIL, Saint-Aubin, Gif-sur-Yvette Cedex, France) using the radiation emitted by a bending magnet source and a new generation undulator providing an energy of 8 keV (resolution close to 2 eV). The organoclays were introduced in an house infiltration cell. The available scattering vector q range was between $1.92.10^{-3}$ $\text{Å}^{-1}$ and $0.7548$ $\text{Å}^{-1}$. This scattering vector q is calculated by the eq (2):

$$q = \frac{4\pi \sin\theta}{\lambda} \quad (2)$$

where $2\theta$ is the scattering angle and $\lambda$ is the wavelength ($\lambda=1.418$ Å). Several successive frames of 2s each were recorded for each sample. A check for radiation damage or sample evolution was performed and no difference between frames was found unless otherwise specified. Each frame-averaged scattering spectrum was corrected for the detector response and scaled to the transmitted intensity, using the scattering intensity from a reference glassy carbon sample integrated over a given angular range. Standard deviations of each measurement were computed as the square root of the number of detected photons.

# Results

**Adsorption isotherms**

Adsorption isotherms highlight the affinity of fluids for porous materials. The good affinity of $C_{10}E_3$ for Na-smectite is shown by the adsorption isotherm in Figure 2. In the whole studied concentration range, the isotherm displays an increase of the adsorbed amount of nonionic surfactant with the equilibrium concentration of $C_{10}E_3$ in solution is added. No plateau is observed and adsorbed amount of $C_{10}E_3$ shows stronger values than previous studies performed on the adsorption of both nonionic and cationic surfactants on clay minerals.[8,9,21,23] Indeed, in previous studies, the adsorption of both cationic and nonionic surfactants on clay is described by concave isotherms fitted by the model of Langmuir where the adsorbed amount approaches a plateau for high loadings of surfactants. In the case of cationic surfactant, the adsorption process is well described in the literature and is explained in two steps. The first stage of the adsorption corresponds to an ion exchange in stoechiometric proportion between surfactant cations and inorganic exchangeable cations which are expelled from the interlayer spaces and thus found in solution.[8-10] Once, the anionic sites of the clay surfaces are neutralized, a new adsorption step occurs by hydrophobic interactions between the alkyl chain of the first adsorbed molecules and the incoming ones. The limit of the organic intercalation depends on the organoclay preparation and is probably due to the saturation of the clay's interlayer space. For nonionic surfactants such as Brij 78 and others polyethylene glycol ether used in previous studies, the adsorption process takes place mainly below but near the CMC and involves a condensation of monomers which leads to a confinement of two molecular layers parallel to the silicate surface.[21,23,24]

Adsorption isotherm of $C_{10}E_3$ on Na-smectite differs from previous studies on several points. First, the isotherm cannot be fitted by a Langmuir model. Second, it is characterized by the absence of any plateau in the studied concentration range. Third, it shows a large amount of adsorbed surfactant as the surfactant concentration increases, suggesting an unrestricted monolayer-multilayer adsorption. This is probably due to the sample preparation where organoclays were realized above the CMC of the $C_{10}E_3$ in lamellar phase. Thus, the adsorption mechanism should occur rather in aggregates than in the monomer state.[23,25,26] Moreover, the synthesis was achieved in solution where the Na-smectite suspension is well swollen and shows a stable hydration state in which silicate layers are widely expanded and already opened when $C_{10}E_3$ penetrate, thus allowing an easier adsorption.

**Fourier Transform Infrared (FTIR) Spectroscopy**

The FTIR technique provides information on the density of the confined surfactant in clays and more generally on organic compounds confined into porous materials. The FTIR spectra between 2800 and 3000 cm$^{-1}$ for $C_{10}E_3$ lamellar phase, the $C_{10}E_3$-clay composites at different initial surfactant concentrations from $3.3.10^{-3}$ to $1.3.10^{-2}$ mol.L$^{-1}$ and the Na-smectite clay are shown in the Figure 3. The spectra were normalized to the most intense band to the stretching mode Si-O-Si at 1030 cm$^{-1}$. The spectrum of the pure $C_{10}E_3$ was divided by a factor of 3 in order to visualize the intensities of the bands in the same window than organoclays spectra. Spectra displayed a broad band in the range 3000-3600

cm$^{-1}$ (not shown), corresponding to the OH stretching of water, suggesting the presence of a small amount of adsorbed water. The intensities of the two intense absorption bands at 2924 and 2852 cm$^{-1}$, corresponding to the asymmetric and symmetric CH$_2$ stretching mode respectively, gradually rise with the increase of the concentration of C$_{10}$E$_3$ within the interlayer space. The intensities of CH$_2$ stretching bands strongly depend on the density of the alkyl chains within the interlayer space.[6,11,27,28] Thus, the FTIR technique can be used as a complementary technique of adsorption isotherms and gives a qualitative idea on the adsorption process.[9]

The FTIR technique is a useful tool to characterize the degree of molecular order of the surfactant in organoclays. This is because the surfactant component consists of long alkyl chains, where the frequencies of CH$_2$-stretching bands can reflect the state of conformational order in the adsorbed organic compounds. For a completely disordered structure, such as alkanes, the characteristics frequencies are 2930 cm$^{-1}$ for CH$_2$ asymmetric stretching and 2853 cm$^{-1}$ for CH$_2$ symmetric stretching.[6,11,27-29] For well-ordered layers, such as crystalline paraffins, the characteristic frequencies are 2916-20 cm$^{-1}$ for CH$_2$ asymmetric stretching and 2846-50 cm$^{-1}$ for CH$_2$ symmetric stretching. For C$_{10}$E$_3$ lamellar phase, these bands appear at 2849 and 2918 cm$^{-1}$, underlining alkyl chains adopt an essentially all-trans conformation. When confined into clays, the wave numbers of both asymmetric and symmetric CH$_2$ stretching modes of C$_{10}$E$_3$ shift to higher frequencies, indicating the introduction of small conformational disorder in the alkyl chains which depends on the concentration of C$_{10}$E$_3$ which can be seen in Figure 4. For a high concentration range from $6.5.10^{-3}$ to $1.3.10^{-2}$ mol.L$^{-1}$, the wave numbers of both asymmetric and symmetric absorption bands for the confined surfactant keeps relatively constant. They are very close to the frequencies of pure C$_{10}$E$_3$ lamellar phase, which are shown by the solid triangle, in good agreement with previous observations led on intercalation of cationic surfactant in clay. However, different results were obtained by Li and Ishida who studied the confinement of hexadecylamine in montmorillonite where wave numbers of CH$_2$ stretching bands were equal to those found in pure amine.[27-28] In our study, the difference is tiny, suggesting the confined $C_{10}E_3$ adopt mainly an all-trans conformation, which can be explained by the fact that the confined surfactant does not have the same three dimensional order than the bulk lamellar phase displays. In low surfactant concentration range, the frequencies weakly shift towards higher wave numbers, but are still reasonably close to the values of the bulk surfactant indicating a larger number of gauche conformer are created and the alkyl tails of confined $C_{10}E_3$ are poorly packed within the silicate layers.

**Small Angle X-Ray Scattering (SAXS)**

SAXS measurements represent a powerful way to understand the changes in the structure of the clay environment, since the interlayer can be estimated by the measuring the $d_{001}$ spacing. Figure 5 exhibits the SAXS profiles collected at the five concentrations for the organoclays. The resulting patterns allow an accurate study of the Na-smectite interlayer evolution after its adsorption by $C_{10}E_3$ in the lamellar phase. The SAXS of the untreated clay is also shown as a reference in Figure 5 and displays a broad diffraction peak located around q=0.53 Å$^{-1}$, corresponding to an interbasal spacing close to 12 Å. Computer simulation and previous X-ray diffraction experiments showed the swelling of Na-montmorillonite, below the osmotic pressure, corresponds to four stable states at basal spacings of 9.7, 12.0, 15.5 and 18.3 Å due to the hydration of one, two, three and four layers of water around the exchangeable sodium cations, respectively.[30,31] The value of 12 Å demonstrates that the Na-smectite is hydrated of one monolayer and confirms FTIR study which highlighted the hydration of untreated clay.

SAXS spectra of the organoclays, although the signal is masked by the strong scattering of the clay platelets at low q values, show several diffraction peaks at the q located at 0.163, 0.34, and 0.485 Å$^{-1}$. It is interesting to remark that q values of the diffraction peaks correspond to $q_2 \sim 2q_1$ and $q_3 \sim 3q_1$, which reflect the existence of a well ordering in the silicate layers and last order diffraction peak at q located around 0.485 Å$^{-1}$ is only observed for the high concentration range. This demonstrates by the use of the FTIR and SAXS techniques that the intercalation of nonionic surfactant is stable and the $C_{10}E_3$ forms ordered structures within the confined space of the silicate layers. The insertion of $C_{10}E_3$ in the interlayer space expands the distance between the layers, giving a $d_{001}$ spacing of ~38 Å, which corresponds to an intersheet separation of δ = 38 – 9.7 = 28.3 Å, where 9.7 Å represents the thickness of the clay layer.[30,31] This value implies that the inserted surfactant molecules must adopt a perpendicular position between the clay surfaces. More interestingly, the intersheet separation value is very close to this-one of the 27.6 Å bilayer thickness of the $C_{10}E_3$ lamellar phase.[15] Thus, the FTIR study and SAXS measurements suggest that confined $C_{10}E_3$ must adopt a well ordering structure close to this-one of the pure lamellar phase, similar to a surfactant bilayer, in the interlayer space, as paraffin organization can present. Another piece of important information derived from X-Ray Diffraction pattern of organoclay (not shown) was the absence of reflection peaks of the $C_{10}E_3$ in the 2θ region 2-64°. This indicates that $C_{10}E_3$ molecules were not aggregated on the external clay surfaces and were mainly intercalated in the interlayer space.

The Figure 6 shows the d spacing of the silicate layers as a function of the surfactant concentration obtained from SAXS measurements. On the whole studied concentration range, a d spacing increment of ~38 Å is observed and indicates an orientation of the confined surfactant bilayer parallel to the clay surface, whose structure must be similar to paraffin type. For high concentration regime, from $6.5.10^{-3}$ to $1.2.10^{-2}$ mol.$L^{-1}$, diffraction signals from three orders of silicate layers are observed and highlight the well structural organization of the silicate layers. On the other hand, in the low concentration regime, only two diffraction orders of silicate layers are displayed. This is in agreement with the FTIR study where shifts in the wave numbers of $CH_2$ stretching bands were observed and more important than in the high concentration regime. The assembly of the $C_{10}E_3$ structure must be less dense and gives more freedom to confined molecules which show light gauche conform defects in the alky tail. However, the first diffraction order remains at the same q value close to 0.163 $Å^{-1}$ than for high concentration regime which suggests that paraffin type structure of the confined molecules is kept.

## Discussion

The manufacture of the organoclays was achieved when the $C_{10}E_3$ displays a lamellar phase and the SAXS and FTIR spectra support the view that organoclay consists of alternatively stacked units of a lamella bilayer sandwiched by clay platelets. The adsorption of $C_{10}E_3$ on clays strongly differs from previous observations on nonionic surfactants where the adsorption resulted to a monomer condensation and molecules lied flat on the silicate surface as monolayer or bilayer. The adsorption of nonionic surfactants is driven by two main chemical interactions: hydrogen bonds and ion-dipole interaction, and still remains a subject of debate.[20,21] These interaction modes are on the same order of magnitude, close to 20 kJ.mol$^{-1}$ and H-bonds involved nonionic surfactants are even stronger than ion-dipole interaction. Moreover, such amphiphilic molecules are typical polar molecules which have displayed their ability to create easily H-bonds with surface silanols groups of either heterogeneous surface and mesoporous materials.[23-26] In previous experiments, Parfitt et al. studied the adsorption of uncharged organic polymers by various exchanged montmorillonites and showed that direct interaction between the exchangeable ions and nonpolar molecules is not responsible for the adsorption, because the cations retain their hydration shells.[20] On the another hand, in a complementary XRD and FTIR study, Deng et al. proved that the adsorption of nonionic surfactants was achieved by an ion-dipole interaction[32].

In such a case of weak interactions process, Cases et al. proposed a model to represent the adsorption of nonionic surfactants on heterogeneous surface.[26] In their model, the authors pointed out the fact the adsorption process occurs from an aggregation mechanism which can be of two or three dimensions type and depends on the surface heterogeneity, the existence of lateral and normal bonds, and the interaction energy involved between the surface and the surfactant. In addition, the adsorption process mainly takes place below but near the CMC and involves a surface aggregative process similar in some respect to the bulk micellization occurring in solution above the CMC. From low to high coverage of the surface, it leads to a local and semilocal structure of adsorbed phase which appear to be "fragmented" bilayer.[23,26] The synthesis of the organoclays was realized above the CMC, in a lamellar phase in which the cohesion of the molecules is reinforced by lateral hydrophobic interactions. The clay minerals present in their structure exchangeable cations equilibrating the negatively charge of the silicate layers. In solution, the presence of water surrounding the exchangeable sodium cations amplifies the repulsive forces at long range order which compete to the adsorption mechanism of the $C_{10}E_3$. Nevertheless, the planar geometry of the platelets and the resulting aggregation mechanism on the silicate surface implies the condensation of the nonionic surfactant in it bulk phase state and explains the bilayer packing of $C_{10}E_3$ within the interlayer space. Moreover, the last observations on POSS cationic surfactants where a large d-spacings was achieved at low surfactant loadings and was attributed to the bilayer packing induced by the bulkiness of the surfactants molecules.[12-14] This underlines the importance of a bulk lyotropic liquid crystal phase state for the organoclay synthesis, which affect the organization of the confined surfactant molecules within the interlayer space.

The density of the $C_{10}E_3$ is approximately 0.94 g.cm$^{-3}$ and its molecular volume is 0.50 nm$^3$, which yields an apparent packing area per molecule of 0.76 nm$^2$. The specific surface area of a Na-montmorillonite is close to 720 m$^2$.g$^{-1}$. If we assume that this specific surface is totally accessible, the calculated amount of adsorbed $C_{10}E_3$ is 1.74.10$^{-3}$ mol.g$^{-1}$, which represents the highest dense packing of the molecules in bilayer state or paraffin type within the interlayer space. In the low regime of concentration, the adsorbed amount represent 41 and 50 % of the highest dense chain packing of the molecules in bilayer structure within the interlayer space respectively. However, the organoclays show the confinement of a bilayer whose SAXS spectra do not present the same degree of order of the silicate layer than for high concentration regime. Moreover, FTIR spectra display the existence of small gauche conformer. $C_{10}E_3$ is adsorbed by an aggregation mechanism as Cases et al. suggested, on hydrophilic sites of the silicate platelets whose planar geometry orientates and orders the lamellar phase.[26] However, the whole clay area is not covered for the low concentration regime, and the confined molecules must own more freedom and a stronger mobility of the alkyl chains which increase consequently the disorder and gauche conformer and thus explains the small observed frequency shift.[27,28] Then, for the another concentration regime, the confined molecules are self organized also in a bilayer within the interlayer space but present a more ordered trans conformer and denser packing structure to minimize the energy and cover until 80% of the surface than the densest bilayer should covered. The electrostatic repulsive silicate surface forces and the no complete covering of the area of the clay platelets may be some explanations for the persistent shift for the high concentration and why we do not recover the same value for the wavenumbers of the $C_{10}E_3$ lamellar phase.

## Conclusion

A main part of the synthesis of organoclays is achieved by ion-exchange with the exchangeable cations in the interlayer space. By this way, cationic surfactant or other particular ionic organic molecules can be intercalated in the structure of the clays. Another way of synthesis proposed for the manufacture of organoclays is the use of nonionic surfactants which allow keeping the charge characteristics of the clay unchanged. However, the adsorption resulted to the condensation in monomeric state and leaded to the intercalation of nonionic surfactant molecules lying on the silicate surface on monolayer or bilayer. The introduction of the $C_{10}E_3$ lyotropic liquid crystal which shows a lamellar phase above the CMC allows a three dimensional aggregation mechanism on the platelets whose planar geometry orders the adsorbed aggregates.

We showed that the adsorption of $C_{10}E_3$ in the lamellar phase in the interlayer space of a Na-montmorillonite by adsorption isotherm, FTIR and SAXS experiments. The behavior of the confined $C_{10}E_3$ differs from previous studies on nonionic surfactants. The confined $C_{10}E_3$ present a bilayer structure within the interlayer space, resulting to the state of the lyotropic liquid crystal in a lamellar phase. Two main features are clear: the lateral hydrophobic interactions between the alkyl chains contributing to the cohesion of the bilayer are stronger than the electrostatic repulsive forces of the silicate surface and the planar geometry of the silicate layers which orders the lamella of the confined $C_{10}E_3$.

The confinement of a bilayer of nonionic surfactant within the interlayer space drives to a hydrophobic organoclay. Moreover, the manufactured organoclay shows a high expansion of the interlayer space and the surface keeps its charge characteristics, i.e., no ion exchange is performed. Thus, this organoclay opens the way to new perspectives for the realization of hybrid nanomaterials. Indeed, cation exchanges can be undertaken, allowing the intercalation into the $C_{10}E_3$-clay structure further functional organic molecules of important size, due to the maximum opening of the basal spacing and the hydrophobicity of the hybrid material.

## Acknowledgments

We thank F. Villieras and C. P. Royall for fruitful discussions. SAXS Experiments were performed at the SOLEIL Synchrotron Facility (CNRS, Saint-Aubin, France). The assistance of F. Meneau during this project was particularly appreciated.

# Figure captions

Figure 1: Phase equilibria of the $C_{10}E_3$-$H_2O$ system. The phases shown are the lamellar ($L_\alpha$), reverse micellar ($L_2$), and sponge ($L_3$) phases. The inserts represent schematic pictures of the lamellar and sponge phases and the structures of the phase were observed by phase contrast microscopy. (adapted from[15])

Figure 2: Adsorption isotherm of $C_{10}E_3$ in the lamellar phase (the critic micellar concentration is $6.10^{-4}$ mol.L$^{-1}$) onto a Na-smectite clay from Wyoming. The black line is guide for the eyes.

Figure 3: Progressive growth of the density of the alkyl chains within the interlayer space of the silicate layers. The sample initial surfactant concentrations are (b) $3.3.10^{-3}$ mol.L$^{-1}$, (c) $4.9.10^{-3}$ mol.L$^{-1}$, (d) $6.5.10^{-3}$ mol.L$^{-1}$, (e) $9.8.10^{-3}$ mol.L$^{-1}$, and (f) $1.3.10^{-2}$ mol.L$^{-1}$. The symmetric and asymmetric $CH_2$ stretching whose wave numbers indicate the presence of disorder / order in the alkyl chains, correspond to the two bands. Untreated Na-smectite clay (a) and pure $C_{10}E_3$ in lamellar phase (g) also are represented. The last spectrum is affected by a ratio (divided by 3) for a visibility in the same window scale. The spectra resolution is 1 cm$^{-1}$.

Figure 4: Representation of the change in frequency of the symmetric and asymmetric $CH_2$ stretching of the five organoclays as a function of the starting $C_{10}E_3$ concentration.

Figure 5: Small Angle X-Ray Scattering (SAXS) profiles for the organoclays at the initial surfactant concentration from (b) $3.3.10^{-3}$ mol.L$^-$1 to (f) $1.3.10^{-2}$ mol.L$^{-1}$. The untreated Na-smectite (a) is added as a reference.

Figure 6: The d spacings of 001 silicate layers obtained from SAXS profiles as a function of the initial $C_{10}E_3$ concentration.

# Figures

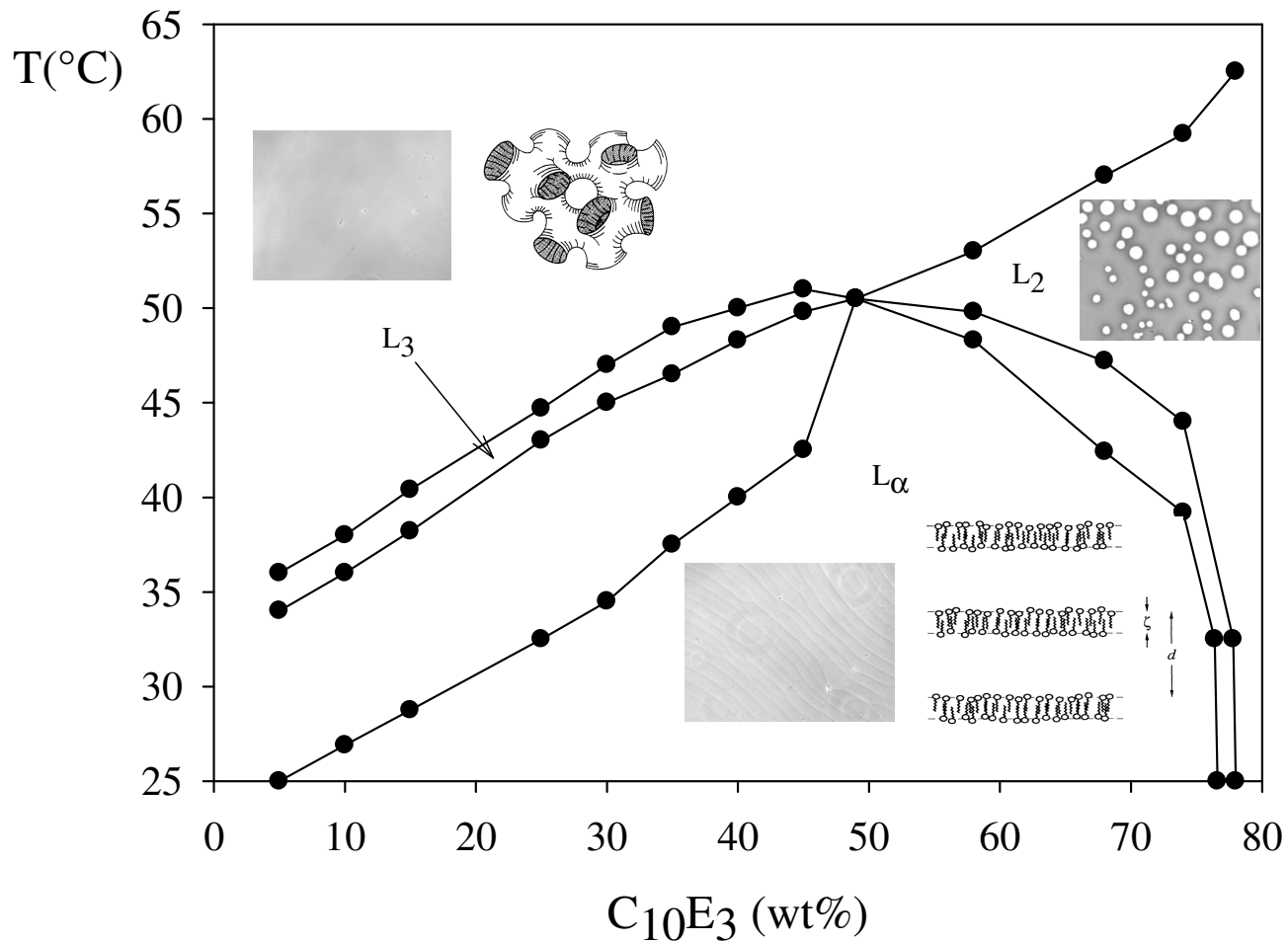

Figure 1: Intercalation of a Nonionic Surfactant ($C_{10}E_3$) bilayer into a Na-Montmorillonite Clay

R. Guégan et al.

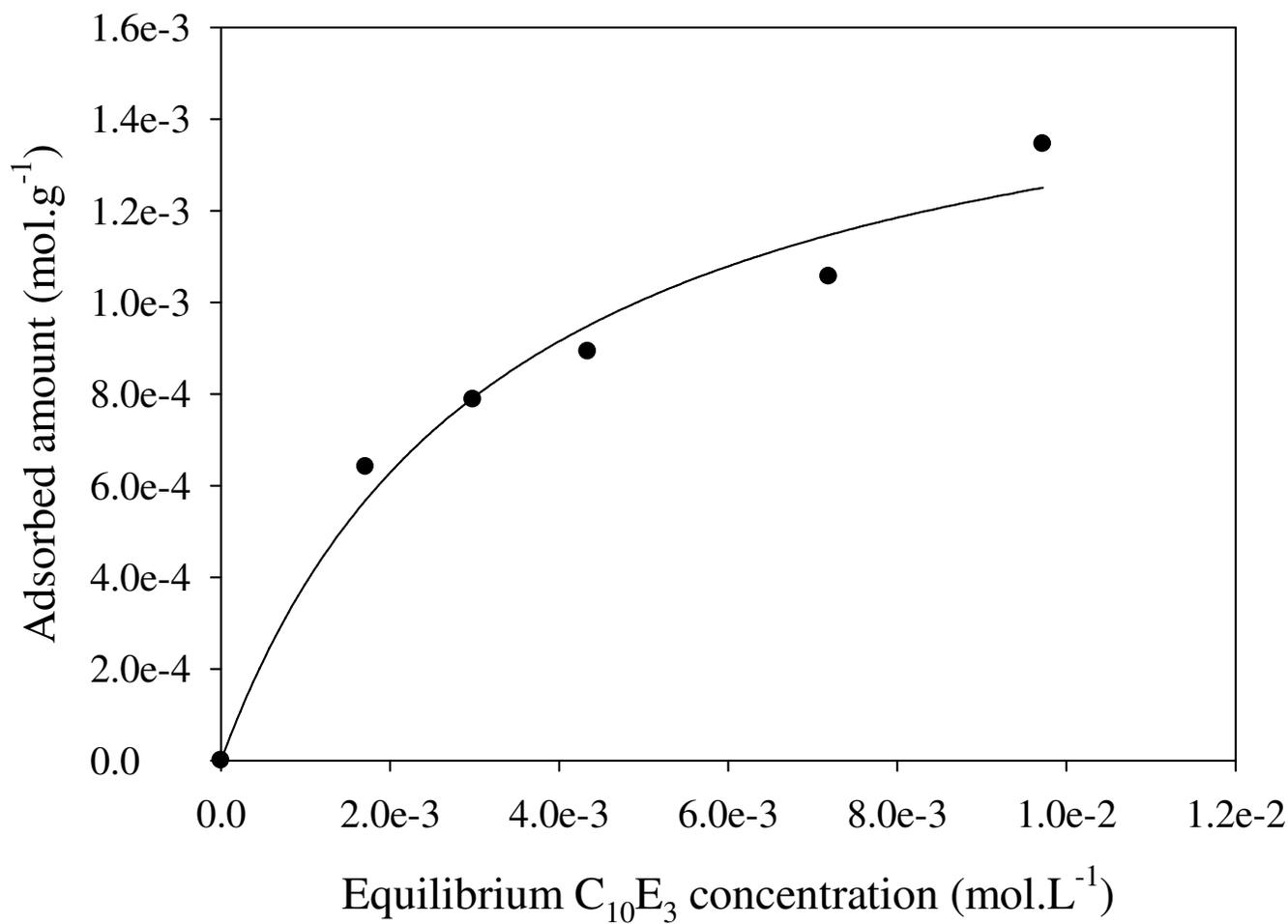

Figure 2: Intercalation of a Nonionic Surfactant ($C_{10}E_3$) bilayer into a Na-Montmorillonite Clay

R. Guégan et al.

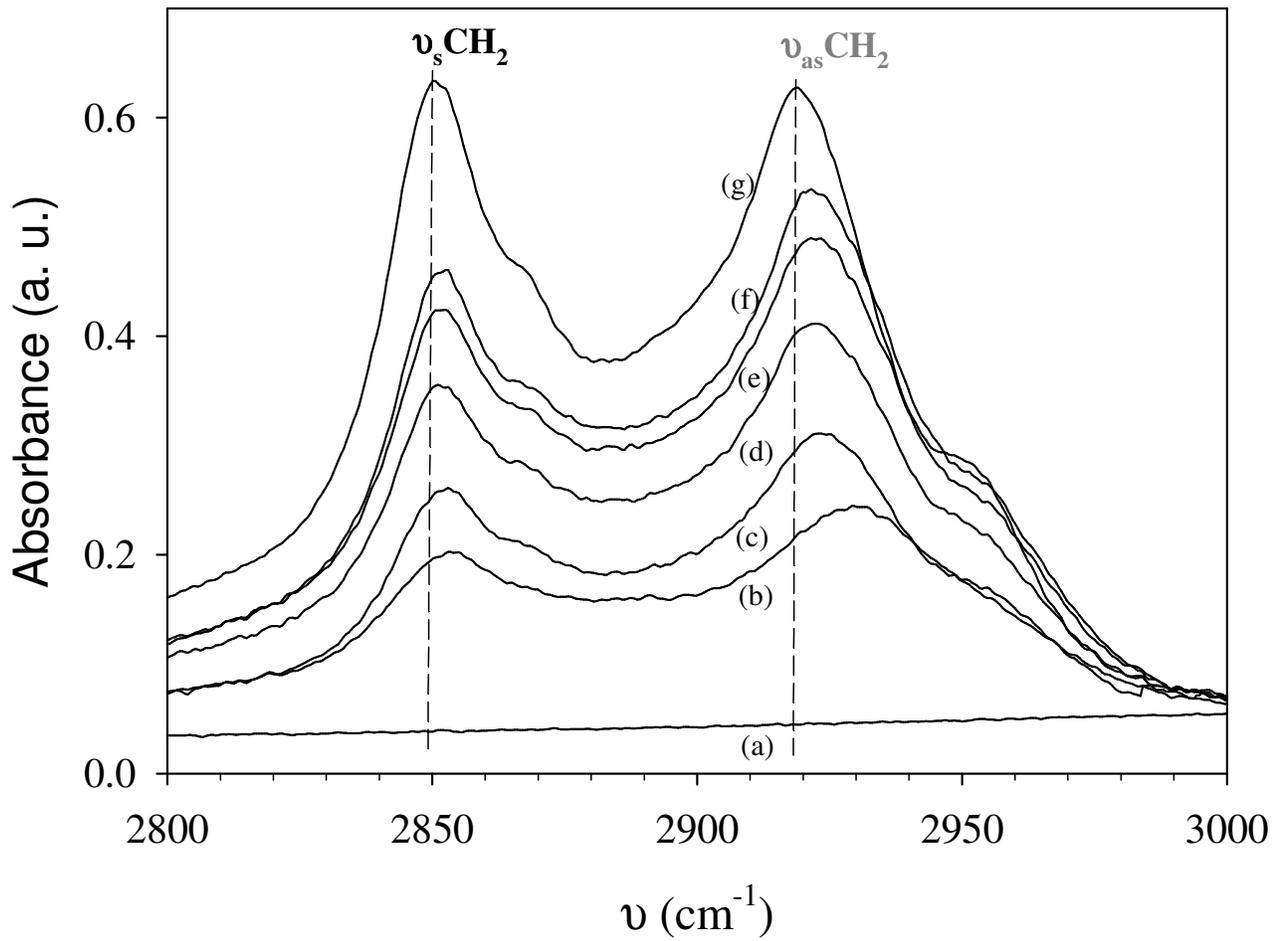

Figure 3: Intercalation of a Nonionic Surfactant ($C_{10}E_3$) bilayer into a Na-Montmorillonite Clay

R. Guégan et al.

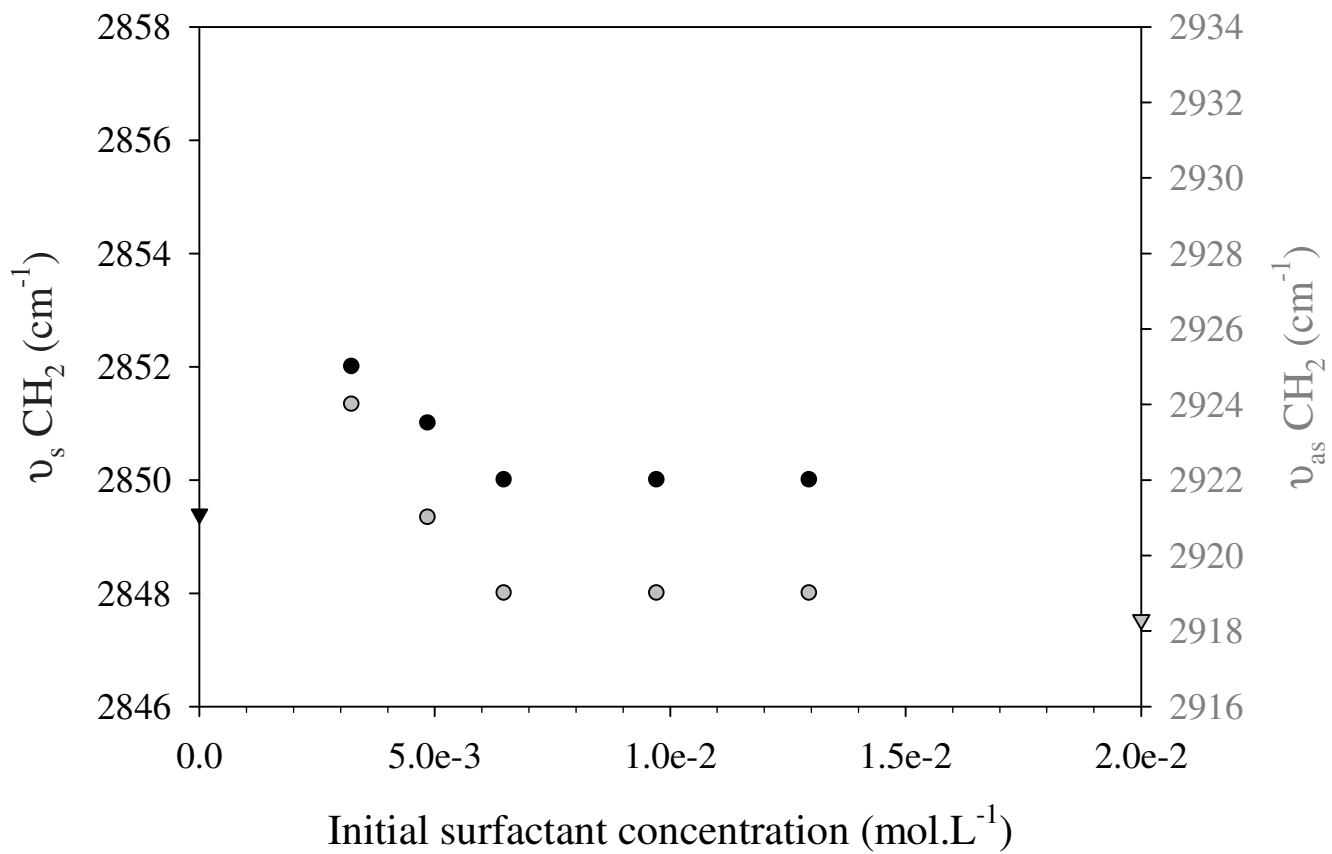

Figure 4: Intercalation of a Nonionic Surfactant ($C_{10}E_3$) bilayer into a Na-Montmorillonite Clay

R. Guégan et al.

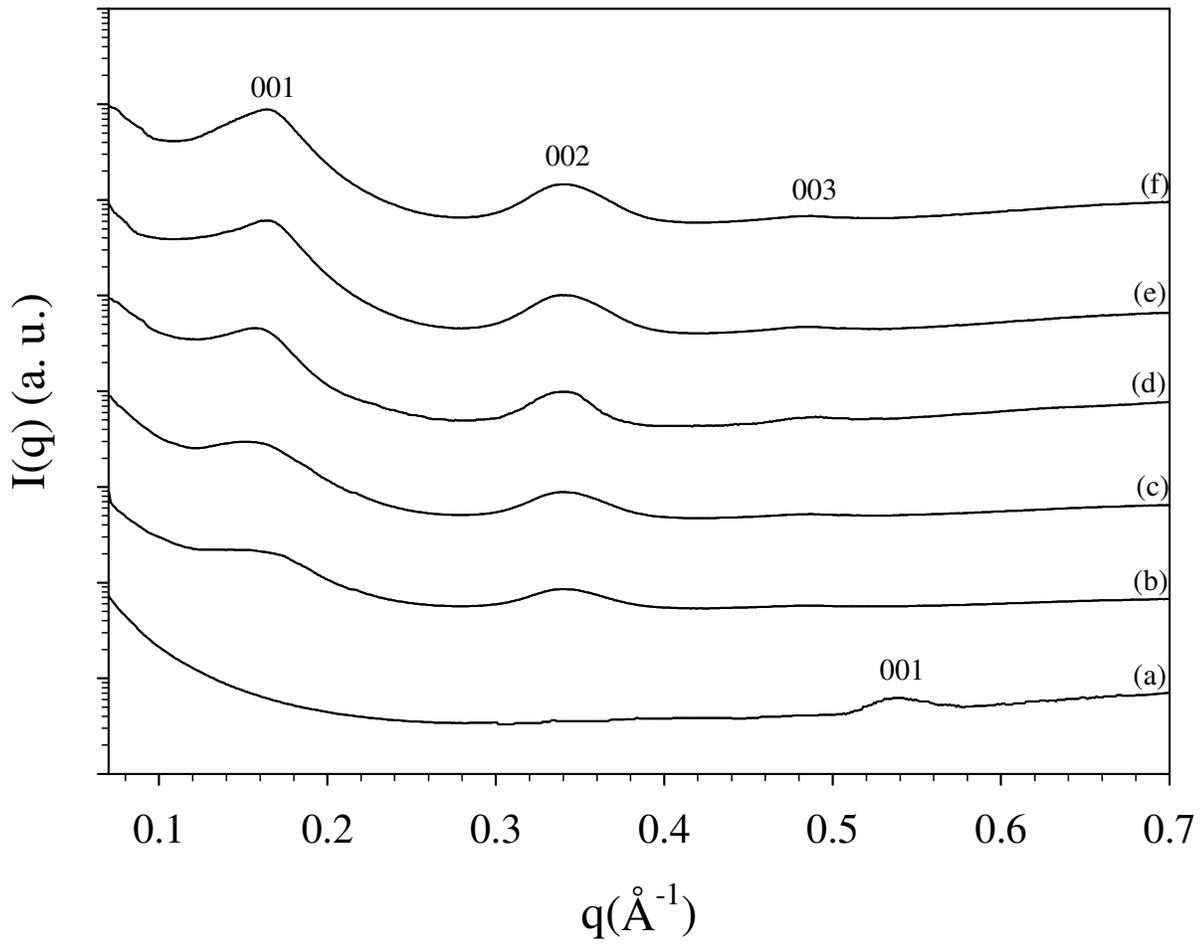

Figure 5: Intercalation of a Nonionic Surfactant ($C_{10}E_3$) bilayer into a Na-Montmorillonite Clay

R. Guégan et al.

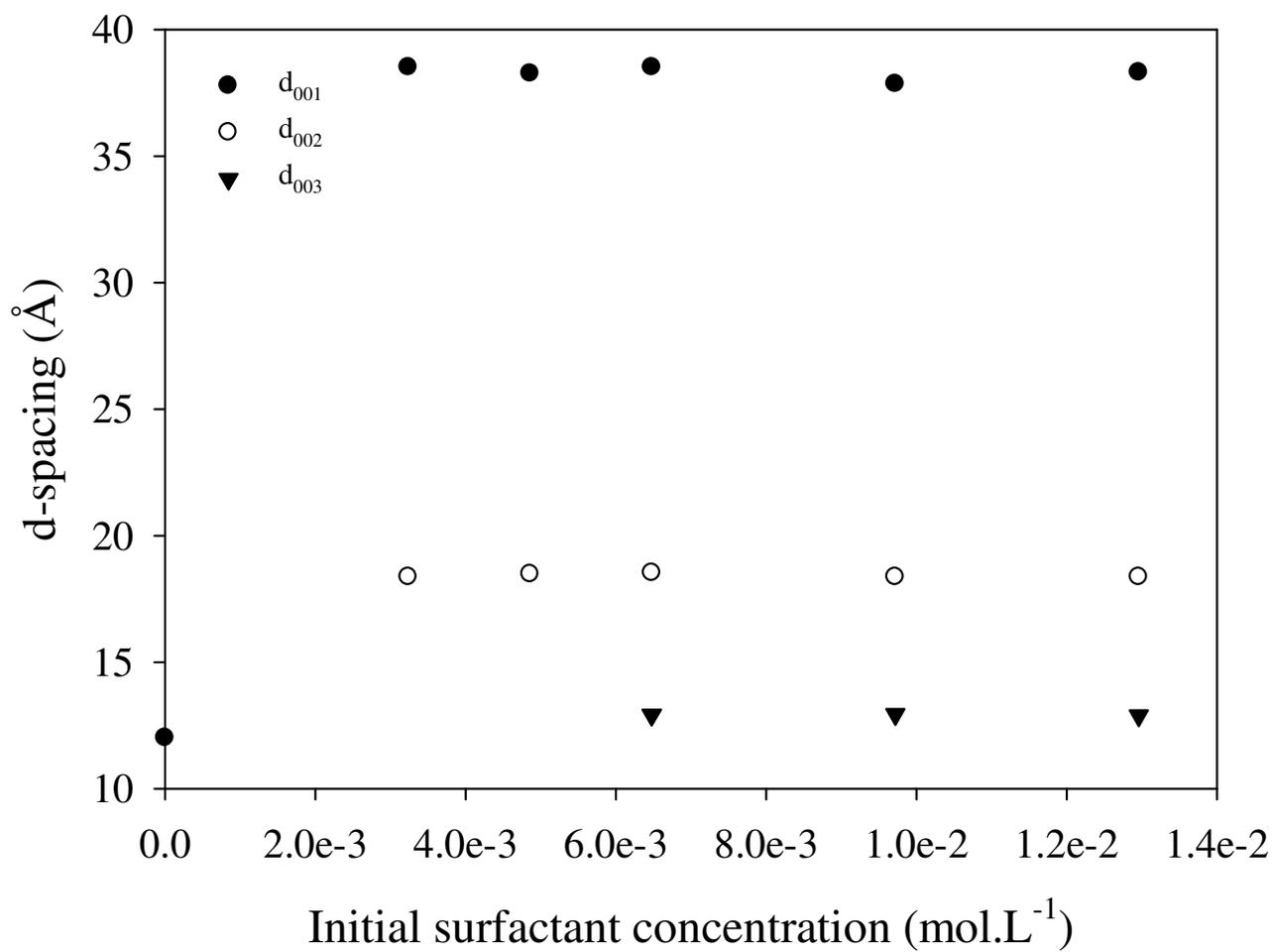

Figure 6: Intercalation of a Nonionic Surfactant ($C_{10}E_3$) bilayer into a Na-Montmorillonite Clay

R. Guégan et al.